body of paper

\documentstyle[preprint,prl,aps,epsfig]{revtex}
\tighten
\begin{document}
\preprint{\today}
\draft
%
%
\title{Ultrashort Laser-pulse Diagnostics for Detection\\
of Ordering within an Ion beam.}
\author{R. Calabrese{\dag}, V. Guidi{\dag}, P.Lenisa{\dag},
 E. Mariotti{\S}, L. Moi{\S} and
 U.Tambini{\dag}}
\address{\dag Dipartimento di Fisica dell'Universit\`a, I-44100 Ferrara, Italy;
	INFN - Sezione di Ferrara, Italy.\\
\S Dipartimento di Fisica dell'Universit\`a, I-53100 Siena, Italy.}
\date{\today}
\maketitle
%
%
\begin{abstract}

A novel diagnostic method to detect ordering within
one-dimensional ion beams in a storage ring is presented. The
ions are simultaneously excited by a ultrashort pulsed laser ($\simeq 1$ ps)
at two different locations along the beam and fluorescence
is detected by a group of four photo-multipliers.
Correlation in fluorescence signals is a
firm indication that the ion
beam has an ordered structure.
\end{abstract}
%
%
%
%
\narrowtext
\section{Introduction}
The possibility to observe ordered structures within an ion
beam has met the interest of the community of accelerator people
\cite{[4]}. Afterwards an even bigger effort has been put in researching
on this new topic \cite{work}. Molecular dynamics approaches
\cite{schif,hasse,schuck} and theoretical investigations \cite{sessler}
have studied in some detail what one could achieve in a storage ring.
A feasibility study for a storage ring dedicated to crystallization has
been recently carried out\cite{lnl}.
As far as present knowledge is concerned no direct evidence of an ordered
structure of an ion beam in a storage ring has been recorded.
Some experiments \cite{russi,gsi} showed that, under particular circumstances,
 suppression of intra-beam scattering occurred; therefore the seek for a
direct and unambiguous way to detect ordering within an ion beam is a
crucial question.

Experiments carried out in ion traps have shown that a record of
crystallization is achieved by observing the fluorescence signal of
ions through a CCD camera \cite{[2]}. Unfortunately this technique cannot be
applied straight away to accelerators since ions are traveling at
relatively high velocity.
 A novel method to detect ordering of a one-dimensional ion beam by resolving
 the single ion structure has been recently proposed\cite{nostro}.
The technique is based on pulsed-laser excitation of the ions in the beam.
In this paper we analyze the electromagnetic-radiation--ion interaction
in order to demonstrate the feasibility and the non-invasivity of the method.
A specifically devoted Monte Carlo simulation will show the results of a
possible application of the method to two of the existing storage rings.

\section{The diagnostic device }
\label{dev}
We consider a pulsed laser with frequency $\nu = (E_2 - E_1)/h = \nu_{ion}$,
where $\nu_{ion}$ is the frequency of an allowed transition between the energy
levels of an ion in the beam.
The laser beam is split into two parts,
which simultaneously cross the ion beam at right angles at two nearby
positions along the storage ring (see Fig.1).
This laser-to-ion crossing area is followed by four photo-multipliers
which detect the photons emitted by the ions that have previously been
excited by the laser beams.
The signals recorded by the photo-multipliers are analyzed when
one laser beam is moved with respect to the other one. In the absence of
ordering, no correlation in fluorescence signal should be
recorded while changing the relative distance of the two laser beams.
 On the contrary, if a string were obtained as a result
of cooling then a strong correlation between the signals should be
observed.
Suppose that one of the four photo-multipliers detects the fluorescence
of an ion excited by one of the laser beams. If one of the other three
photo-multipliers detects a simultaneous fluorescence signal, it means that
the other laser has interacted with another ion in the string, in turn
indicating that
the distance between the two ion-to-laser crossing points is an
integer multiple of the inter-particle spacing of the string.
 Then, by slightly
moving the second laser beam, the correlation signal should vanish.
A sort of periodical dependence on the distance between the laser beams
should appear.
The above considerations were related to a perfect string of ions. At non--zero
temperature ions oscillate along both longitudinal and transversal directions.
Choosing the distance between the laser beams of the order of
interparticle spacing the effect of longitudinal oscillations is negligible
as the distance between two nearest-neighbor particles can be regarded to be
relatively uniform for a plasma parameter of the order of 100 \cite{nostro}.
If the laser beams are focused on the ion string through cylindrical lens
also the effect of transverse oscillations becomes negligible.
The particles in the ion beam travel
with velocity $v=\beta c$; if $s$ denotes the interparticle distance then the
time taken by an ion in the beam to travel this distance
is $T=s/v$.
Typical interparticle distance for a string in a storage ring is
$s=10-100~\mu$m;
 considering a velocity range useful for existing storage rings, i.e. $\beta =
0.003 - 0.05$, $T$ lies from 1 to 50 ps.
The diagnostic method is aimed at revealing the single-particle ion
structure using laser pulses which are shorter than the time $T$.
In this way the ions can be regarded to be at rest and all the problems
connected
with analyzing a traveling structure are overcome.
Thus, the laser pulse duration $\Delta t$ must be of the order of one
picosecond or fraction of ps.

An electromagnetic radiation with this very short temporal duration
is far from being monoenergetic and  this has to  be taken
into account in order to
 show the feasibility of the method.

\section{Laser pulse -- ion beam interaction}
\label{int}
The electric field  of the laser
 is given by the following expression \cite{adams}
\begin{equation}
{\bf E}(t)= {\bf e(r)} {\cal E} (t) \{ a_L e^{i \omega_0 t} +
a_L^{+} e^{- i \omega_0 t} \}
\end{equation}
where the gaussian shape for ${\cal E} (t)$
\begin{equation}
{\cal E}(t) = \frac{E}{2} e^{-\frac{(t-t_0)^2}{2 \sigma^2}}
\end{equation}
is assumed to account for the finite duration of the laser pulse.


Considering the time duration and the frequency bandwidth
of the intensities  at FWHM\cite{Shapiro}
\begin{equation}
\Delta t / 2 = \sigma  \sqrt{\ln2};\qquad\Delta \omega/2=\sqrt{\ln2}/\sigma
\label{delta}
\end{equation}
the time-bandwidth product gives
\begin{equation}
\Delta t \cdot \Delta \nu = \Delta t \cdot \frac{\Delta \omega}{2 \pi}=.441
\label{allargamento}
\end{equation}
This means that a laser pulse of the order of one ps has a frequency bandwidth
of
hundreds of GHz, i.e.,  is far from being monochromatic.
This bandwidth is larger than hyperfine and fine splittings of the ions that
can be easily laser cooled, i.e. best candidates for reaching
crystallization.
In the following we shall neglect the details of subtransitions, therefore
laser-ion interaction will be described as a two-level quantum mechanical
system with an external time-dependent perturbation \cite{QM}($\vert
 g \rangle$, $\vert e \rangle$ being ground and excited states).

The Hamiltonian of the system is:
\begin{equation}
H=H_0+V_I
\end{equation}
where the external potential is
\begin{equation}
V_I=- {\bf d} \cdot {\bf E} = - d_eg {\bf e}_z \cdot {\bf e (r)} E(t)
\end{equation}
and ${\bf d}=d_{eg} {\bf e}_z$ is the dipole moment for the transition under
consideration.

Assuming the atomic system to be initially in the ground
state, one can calculate the excitation probability $P_2(t)=
\vert c_2(t) \vert^2$ solving the
following differential equation for $c_2(t)$

\begin{equation}
\ddot{c_2}(t)+ \left[i \Delta \omega +\frac{(t- t_0)}{\sigma^2}\right]
\dot{c_2}(t)+ \Omega^2(t)c_2(t) = 0
\label{diffeq}
\end{equation}
with initial conditions
\begin{equation}
\left\{
\begin{array}{ccc}
c_2(0)&=&0\\
\dot{c}_2(0)&=&-i \Omega(0)
\end{array}
\right.
\end{equation}
where
\begin{equation}
\Omega(t)=\frac{R d_{eg} E}{2 \hbar} e^{-\frac{(t-t_0)^2}{2 \sigma^2}}
\qquad
\Delta \omega = \omega + \omega_{12}.
\end{equation}
and $\omega_{21}=-\omega_{12}=(E_2-E_1)/\hbar$,
$R={\bf e}_z \cdot {\bf e (r)}$.

Let us seek for an analytical estimation of the probability of excitation due
to the laser pulse ($P_2(t) \to +\infty$) by solving Eq. (\ref{diffeq})
heuristically, i.e. neglecting the dissipation term and averaging out
$\Omega (t)$.
In order to have
a non negligible excitation probability  the time integral of
$\Omega$ must be of the order of unity :
\begin{equation}
\int \Omega(t) dt =
\frac{R d_{eg} E}{2 \hbar} \sqrt{2 \pi } \sigma \simeq 1.
\end{equation}
Taking into account that the ions are randomly oriented
(the average $R^2$ over the solid angle gives a factor $1/3$),
using Eq. (\ref{delta})
and the relation between the dipole momentum and the spontaneous lifetime
($\tau$) of the ion, we have  a rough estimate of the minimum value
for the peak intensity of the laser to cause excitation with appreciable
probability:
\begin{equation}
I \simeq \frac{2 \ln(2) \hbar \omega_0^3 \tau}{ \pi^2 c^2 \Delta t^2}
\end{equation}
Introducing the saturation intensity for the transition considered
\begin{equation}
I_s=\frac{\hbar \omega_0^3}{4 \pi c^2 \tau}
\end{equation}
we obtain  an approximate rule for the requested laser intensity
\begin{equation}
\frac{I}{I_s} \geq k \left(\frac{\tau}{\Delta t}\right)^2
\label{approx}
\end{equation}
where $k$ is a constant of the order of the unity.

In order to test the validity of Eq.(\ref{approx}),
one can numerically solve the differential equation (\ref{diffeq}).
Fig.2 shows a contourplot of the excitation probability as a function
 of the nondimensional parameters $\log(I/I_s)$ and $\log(\Delta t/\tau)$.
The bold line corresponds to the approximate rule of Eq.\ref{approx},
with $k=1$; for that value the excitation probability is close to $1/2$.
One can see that the excitation probability is strongly influenced by
the duration and the intensity  of the laser pulse.
In order to show that the excitation probability is not negligible in practical
cases, we shall refer to two presently being laser-cooled ions in their
respective storage rings,
i.e. $^{24}\!Mg^+$ at ASTRID (Aarhus) and  $^9\!\!Be^+$ at TSR (Heidelberg).
Considering a focusing spot of $5\times 200 \mu m ^2$ and a pulsed laser with
some nJ/pulse with 10 \% jitter in intensity and pulse duration, we obtain two
allowed areas for the probability of exciting ions to the upper level.
Due to the focusing of the laser beam the perturbation strength depends
on the transverse coordinates of the ion interacting with the
electromagnetic radiation. This is the major contribution to the uncertainty
over $\log_{10}(I/I_s)$.
The excitation probability averages $P=0.58$ for  $^9\!\!Be^+$ and
$P=0.51$ for $^{24}\!Mg^+$, i.e. the
diagnostic scheme can work for either ions in their respective storage rings.

\section{Monte Carlo simulation}
After having investigated about the possibility of exciting ions to the upper
level by pulsed lasers with sufficiently high probability, we shall
demonstrate the feasibility of the method. The rate of coincidence events
detected by the group of photomultipliers depends on several parameters such as
the geometry of the system, quantum efficiency of photomultipliers, filtering
of fluorescence photons and the dynamical behavior of the system due
to nonzero temperature of the ions in the beam \cite{nostro}.
 We have developed a Monte Carlo to account for all these effects and
to simulate
 the response of the diagnostic device when is being
applied to existing storage rings.
 The Monte Carlo is limited to the cases of $^9\!\!Be^+$ at TSR (Heidelberg)
 and of $^{24}\!Mg^+$ at ASTRID (Aarhus); some important parameters of these
rings are summarized in Table I. Also included in the table are the main
parameters of a laser suited to diagnostics in each of these rings.

\subsubsection*{$^{24}\!Mg^+$ at ASTRID}
Typical velocity of $^{24}\!Mg^+$ at ASTRID is about $\beta=0.003$;
 this corresponds to a time of about 33 ps taken by an ion in the
beam to travel the interparticle spacing. A 2-ps pulse duration of the
laser was considered for in the simulation.
	Figure 3 illustrates the fluorescence response for a string of
$^{24}\!Mg^+$ ions at $\Gamma=100$ with $s=50 \mu$m as a function of the
distance
between the two laser beams. Appearance of fluorescence peaks is
visible in the figure when the distance between the laser beams is
an integer multiple of the string spacing. Also the pattern for a
disordered beam with the same density as the string is shown; in
this case a totally uncorrelated pattern is achieved and the
fluorescence signal is related only to stray counting of the
photomultipliers.
It is remarkable that a clear firm of ordering in the ion beam can
be achieved with only 1 s acquisition time, i.e. much less than the beam
lifetime in the storage ring.

The same simulation was carried out with the same laser and beam
parameters as for Fig. 3 but with a longer acquisition time (10 s).
Correlation peaks have became more neat due to
a longer interval of counting (Fig. 4).

\subsubsection*{$^9\!\!Be^+$ at TSR}

	The simulation was carried out also for the case of $^9\!\!Be^+$ ion
beam at TSR (Heidelberg) at $\Gamma=100$. Ion velocity is higher for this
storage ring ($\beta=0.05$) and therefore one should resort to still shorter
pulse duration of the laser. A commercially available 200-fs pulsed
laser with some nJ/pulse was considered in the Monte Carlo, the results of
which are shown in Fig. 6. Correlation peaks are shorter, with respect to the
case of $^{24}\!Mg^+$, due to the lower quantum efficiency of photomultipliers.

\section{Perturbation to the ion beam}
\label{perturba}
The proposed method of fluorescence detection makes use of absorption
of a photon of the laser and its subsequent re-emission.
As for any kind of diagnostics, this technique perturbs the system under
consideration; here we provide an estimation of the heating rate of the
ion beam owing to its interaction with the laser.

Suppose the ion suffering interaction with the light to be at rest in a given
 frame; when a photon is absorbed the ion receives the impulse of the photon
and moves in the direction of the laser beam.
Imposing impulse and energy conservations, in the ever valid approximation
$\Delta E \ll  mc^2$, where $\Delta E$ is the transition energy of the ion
and $m$ is the ion's mass, the increase in ion velocity holds:
\begin{equation}
v = \frac{\Delta E}{mc}.
\label{vel}
\end{equation}
Then the ion emits a photon isotropically.
Averaging over several absorption-emission processes, an ion gains a mean
transverse velocity according to Eq.(\ref{vel}).
Thus, each interaction with the laser enhances the transverse kinetic
energy by:
\begin{equation}
        E_K = \frac{mv^2}{2} = \frac{\Delta E^2}{2mc^2}
\end{equation}

This energy is usually referred to as recoil energy.

The energy increase per second of the ion string is $E_K$ times the  rate
of excitations, $f$, to the upper level for each laser beam.
Under the assumption that this energy will be shared between the other
particles
through intra-beam collisions, the rate of temperature increase for the ion
 beam is:
 \begin{equation}
 \frac{dT}{dt} = \frac{2 E_K f}{N k_B}
 \end{equation}

For the case of $N=8\cdot 10^5 \ ^{24}\!Mg^+ (E_K/k_B=2.6 mK)$ ions circulating
in the ASTRID
storage ring and with the above parameters for the diagnostics
($f=2.5\cdot10^6$ Hz), it comes out a rate:
\begin{equation}
     \frac{dT}{dt} = 1.6\cdot10^{-5} K/s
\end{equation}
{}From the above calculation it emerges that the perturbation to the ion
beam due to the diagnostic device is negligible and should not affect
the ordering within the ion beam.
In the calculation was assumed that the energy pumped by the laser to
the ion beam was perfectly shared by all the ions. In reality,
intrabeam scattering of highly cooled beams is expected to be considerably
damped and areas with a local temperature higher than the average one
should appear.
However, the amount of energy given by the laser to a single ion through
excitation is so small that these local temperatures should not affect
significantly the ordering in the beam.

\section{Conclusions}
\label{concl}
It has been shown that the method is useful as a diagnostic
tool to detect ordering in one-dimensional systems. The method
enables to resolve the single-ion structure of the beam. Finally we
point out that all the components of the system are
commercially available equipements and that the diagnostics could be easily
implemented in one of the existing storage rings.

%
%

%

\begin{table}
\begin{tabular}{ccc}
Storage ring &	TSR &	ASTRID\\
$\beta$ &	0.05 &	0.003\\
ion species&	$^9\! Be^+$&	$^{24} \!Mg^+$ \\
wavelength&	313 nm&	280 nm\\
upper state lifetime& 8.7 ns & 3.5 ns\\
lower state lifetime&	groundstate&	groundstate\\
laser repetition rate&	50 MHz&	50 MHz\\
pulse duration&	200 fs&	2 ps\\
bunch energy & 3 nJ & 3 nJ\\
excitation probability & 0.58 & 0.51 \\
\end{tabular}
\caption{Important parameters for the simulation}
\end{table}
%

\begin{figure}
\caption{a) Layout of the experimental set-up.
b) A sketch of the ion-laser interaction region.}
\end{figure}

\begin{figure}
\caption{Contourplot of the excitation probability $P(t \to \infty)$
as a function of
$log_{10}[I/I_s]$ and $log_{10}[\Delta t/\tau]$. The contourlines correspond
to a probability of 0.05, 0.25, 0.5, 0.75, 0.95, respectively. The bold line
corresponds to the approximate rule of Eq.\protect\ref{approx}. The two boxes
represent the allowed "areas" for $^{24}\!Mg^+$  at ASTRID (left) and for
$^9\!Be^+$ at TSR (right).}
\end{figure}
\begin{figure}
\caption{ Pattern of fluorescence for $^{24}\!Mg^+$ ion beam at
	$\Gamma=100$ for 1 s acquisition time.}
\end{figure}

\begin{figure}
\caption{ Pattern of fluorescence for $^{24} \!Mg^+$ ion beam at
	$\Gamma=100$ for 10 s acquisition time.}
\end{figure}

\begin{figure}
\caption{ Pattern of fluorescence for $^9 \! Be^+$ ion beam at $\Gamma=100$
for 1 s acquisition time.}
\end{figure}
\end{document}